\documentclass[onecolumn,notitlepage,pra,showpacs,superscriptaddress,amssymb,amsmath,amsmath]{revtex4-1}
\usepackage{graphicx}
\usepackage{epstopdf}
\usepackage{bm}
\usepackage{hyperref}
\usepackage{cleveref}
\usepackage{comment}
\usepackage{color}
\usepackage{physics}
\usepackage{amsmath}
\usepackage{tikz}
\usepackage{enumitem}
\usepackage{bbold}
\usepackage{ulem}

\hypersetup{%
   pdfpagemode=None, 
   pdfstartpage=1,
   pdfmenubar=true,
   pdftoolbar=true,
   colorlinks = true,
   linkcolor=blue,
   citecolor=blue,
   urlcolor=blue,
   bookmarksopen=false
 }

\def\<#1>{\mathinner{\langle#1\rangle}}

\newcommand{\be}{\begin{equation}}
\newcommand{\ee}{\end{equation}}

\newcommand{\oper}[1]{\mathbf{#1}} 
\newcommand{\opp}[1]{#1} 
\newcommand{\ver}[1]{\hat{#1}}

\newcommand{\beq}{\begin{equation}}
\newcommand{\eeq}{\end{equation}}

\begin{document}

\title{Controlling the dynamics of ultracold polar molecules in optical tweezers}

\author{Marta Sroczy\'{n}ska}
\affiliation{Faculty of Physics,  University of Warsaw, Pasteura 5, 02-093 Warsaw, Poland}

\author{Anna Dawid}
\affiliation{Faculty of Physics,  University of Warsaw, Pasteura 5, 02-093 Warsaw, Poland}
\affiliation{ICFO - Institut de Ci\`encies Fot\`oniques, The Barcelona Institute of Science and Technology, 08860 Castelldefels, Spain}

\author{Micha\l~Tomza}
\affiliation{Faculty of Physics,  University of Warsaw, Pasteura 5, 02-093 Warsaw, Poland}


\author{Zbigniew~Idziaszek}
\affiliation{Faculty of Physics,  University of Warsaw, Pasteura 5, 02-093 Warsaw, Poland}

\author{Tommaso~Calarco}
\affiliation{Forschungszentrum J{\"u}lich, Institute of Quantum Control (PGI-8), 52425 J{\"u}lich, Germany}

\author{Krzysztof~Jachymski}
\email{Krzysztof.Jachymski@fuw.edu.pl}
\affiliation{Faculty of Physics,  University of Warsaw, Pasteura 5, 02-093 Warsaw, Poland}

\date{\today}

\begin{abstract}
Ultracold molecules trapped in optical tweezers show great promise for the implementation of quantum technologies and precision measurements. We study a prototypical scenario where two interacting polar molecules placed in separate traps are controlled using an external electric field. This, for instance, enables a quantum computing scheme in which the rotational structure is used to encode the qubit states. We estimate the typical operation timescales needed for state engineering to be in the range of few microseconds. We further underline the important role of the spatial structure of the two-body states, with the potential for significant gate speedup employing trap-induced resonances.

\end{abstract}

\maketitle

\section{Introduction}
In recent years, ultracold polar molecules have been a subject of intense experimental and theoretical studies. Due to their rich internal structure and comparatively strong intermolecular interactions, they have been suggested as great candidates for quantum simulation~\cite{Gorshkov2011a,Sowinski2012,Doccaj2016} and computation~\cite{DeMille2002,Yelin2006, Ni18, Hughes19,Campbell2020,Albert2020}, precision measurements of fundamental constants~\cite{Krems2009,Safronova2018}, as well as controlled chemistry \cite{Ospelkaus10, Ni10, Liu2021}. The complex internal structure of molecules offers broad prospects for experimental control with external fields~\cite{Micheli2007,Gorshkov2011b,Lemeshko2013,Kruckenhauser2020,DawidPRA18, DawidPCCP20, Caldwell2020}, but also leads to problems with cooling the system to reach quantum degeneracy. These include the lack of suitable cycling transitions (with several notable exceptions), and high inelastic collision rates leading to losses. Nowadays, after a series of experimental breakthroughs, an increasing number of groups can produce large ultracold molecular samples with high phase-space density, as well as trap them in an optical lattice or tweezer array with high filling and low entropy~\cite{Molony2014,Guo2016,Liu2018,Collopy2018,Anderegg2018,Anderegg2019,deMarco2019,He2020,Cairncross2021}. 

While general working principles of quantum engineering with trapped polar molecules are intuitive and based on well-established analogues in atomic systems~\cite{Hayes2007,Kaufman2015,Gross2017,Schafer2020}, to provide meaningful experimental predictions it is essential to account for corrections with respect to commonly used approximations. One crucial aspect is the dipolar interaction, which is not only long-ranged but also state-dependent. As the rotational and hyperfine internal states strongly couple to electromagnetic fields, a suitable choice of molecular species and external conditions can lead to the realization of very diverse and rich many-body models~\cite{Wall2013a,Kruckenhauser2020}. This can be combined with strong optical confinement, which can also be made state-dependent due to the polarizability anisotropy of molecules which determines the trapping frequencies in an optical trap~\cite{Neyenhuis2012}. The characteristic interaction and confinement length scales compete with each other, making tight-binding and pseudopotential approximations questionable. As a result, even for a two-body problem, the full numerical solution requires extensive computational effort~\cite{Buchler,Wall2013,Simoni2016}. Precise information about the structure of states resulting from the interplay of strong interactions and confinement can nevertheless be very beneficial, as it allows to make use of the specific properties of the spectrum to increase the efficiency of state preparation and gate operation. One notable example are the so-called trap-induced resonances~\cite{Stock2006,Doerk2010,Sroczynska2018} resulting from the anticrossing between the molecular-like bound states and the spatially extended trap states. 

In this work, we study the dynamics of a pair of ultracold polar molecules trapped in separate optical tweezers, fully taking into account the trap structure, internal rotational states, dipolar interaction, and external electric field. Our model is generic because it does not rely on any specific feature of particular molecular species and can be described in terms of a few characteristic length and energy scales. We study the system's evolution after a quench of the electric field value and set the stage for future calculations of the dynamics under optimized field pulses, providing estimates for the characteristic time scales.

The paper is structured as follows. In Section~\ref{sec:basic} we introduce the system Hamiltonian and discuss its properties relevant for state engineering. Then in Section~\ref{sec:results} we analyze the spectra and dynamics and introduce the gate protocol, which is further discussed in Sec.~\ref{sec:disc}. Conclusions are drawn in Sec.~\ref{sec:conc}. We provide the code supporting the findings of this study in Ref.~\cite{OurRepo}.

\section{Theoretical model}
\label{sec:basic}
\subsection{System Hamiltonian}
\begin{figure*}[t]
	\begin{center}
		\includegraphics[width=0.8\textwidth]{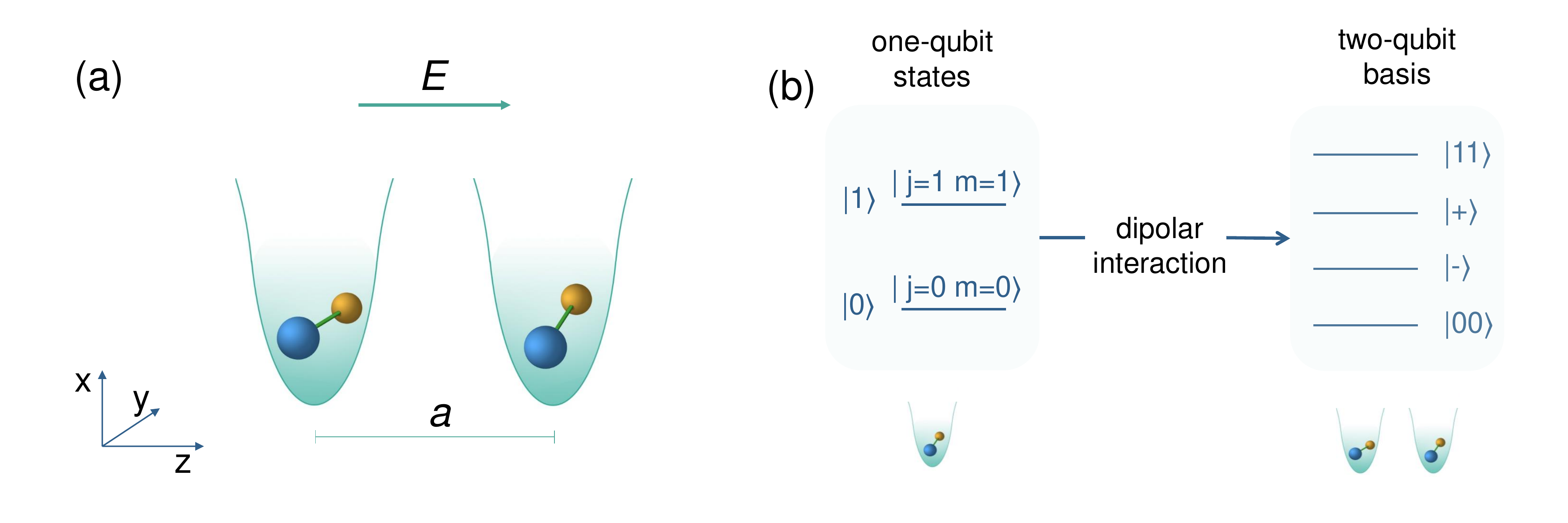}
	\end{center}
	\caption{Schematic representation of the investigated system. (a) Two polar molecules confined in separate one-dimensional traps with electric field applied in the $z$ direction. (b) Qubit states which we analyze in this work. Here $|+\rangle$ and $|-\rangle$ denote denote two-qubit states being symmetric and antisymmetric superpositions $(|10\rangle+|01\rangle)/\sqrt{2}$ and $(|10\rangle-|01\rangle)/\sqrt{2}$, respectively.}
	\label{fig:intro-fig}
\end{figure*}

We consider a system of two polar molecules confined in separate three-dimensional potential wells. For simplicity, we assume the traps to be anisotropic harmonic, but anharmonic corrections are straightforward to include within our approach by extending the basis set to account for the coupled dynamics of the center of mass and relative motion as shown below. The molecules interact with each other via dipole-dipole forces, which are controlled with an external electric field.

Let us begin with a short discussion of the properties of a single polar molecule. Here we solely focus on its rotational structure within the rigid-rotor approximation and assume it remains in the ground electronic and vibrational state due to separation between the internal energy scales. In this work, we also neglect any possible hyperfine structure, as it is not relevant for the current demonstration. 
We choose the $z$ axis along the electric field direction, leading to a simple description of the internal state
\beq\label{Hrot}
H_{\rm rot}=B \oper{j}^{2}-\opp{d}_0 E,
\eeq
where $B$ is the rotational constant, $\oper{j}$ is the rotational angular momentum operator of a single molecule, $E$ denotes the electric field magnitude, and $\opp{d}_0=\ver{e}_0\cdot \oper{d}$ is the $z$ component of the electric dipole moment operator. In general, it is convenient here to use spherical tensor formalism where $\opp{d}_p=\ver{e}_p\cdot \oper{d}=d\, C_p ^{1}(\theta,\phi)$, with $d$ being the value of the permanent dipole moment in the body frame. The functions $C_p^{k}$ denote the unnormalized spherical harmonics $C_p^{k}(\theta,\phi)=\sqrt{4\pi/(2k+1)}Y_p^{k}(\theta,\phi)$, while $\ver{e}_{\pm 1}=\mp (\ver{x}\pm i\ver{y})/\sqrt{2}$ and $\ver{e}_0=\ver{z}$.



When $E=0$, the rotational angular momentum $j$ is a good quantum number. Pure rotational states $\left|j m\right>$ have no mean dipole moment. At nonzero field only the projection of $j$ onto the field axis, $m$, remains conserved and the eigenstates can be decomposed as $\left|\widetilde{jm}\right>=\sum_{j^\prime}{c_{jj^\prime}\left|j^\prime m\right>}$, where $\left|\widetilde{jm}\right>$ denotes the state adiabatically connected to $\left|jm\right>$.  Still, $\opp{j}_z\left|\widetilde{jm}\right>=m\left|\widetilde{jm}\right>$. We will be mainly interested in the lowest lying rotational states connected to $j=0$ and $j=1$. Two natural parameters can be introduced here, the dimensionless field strength $\beta=dE/B$ and the characteristic radius $r_B=\left(d^{2}/B\right)^{1/3}$.


Let us now move to the case of two molecules. The total rotational angular momentum of the system is $\oper{J} = \oper{j}_1 + \oper{j}_2$ with the projection onto the external field axis, $M = m_1 + m_2$. The dipole-dipole interactions between two molecules are commonly represented as
\beq
V_{\rm dd}(\mathbf{r})=\frac{1}{r^{3}}(\oper{d}_1\cdot\oper{d}_2-3(\oper{d}_1\cdot\ver{e}_r)(\ver{e}_r\cdot\oper{d}_2))\, ,
\eeq
where $r\ver{e}_r$ is the vector connecting the two molecules. This expression can also be conveniently rewritten by means of spherical tensors~\cite{Brown}
\beq
V_{\rm dd}=-\frac{\sqrt{6}}{r^{3}}\sum_{p=-2}^{2}{(-1)^{p}C_{-p}^{2}(\theta,\phi)T_p ^{2}(\oper{d}_1,\oper{d}_2)}\, ,
\eeq
which allows for separation of the part conserving the angular momentum projection
\beq
\label{interaction}
\begin{split}
V_{\rm dd}^{p=0}=&-\frac{1}{r^{3}}(3\cos^{2}\theta-1) \left(\opp{d}_0^{(1)}\opp{d}_{0}^{(2)}+\frac{1}{2}\opp{d}_+^{(1)}\opp{d}_{-}^{(2)}+\frac{1}{2}\opp{d}_-^{(1)}\opp{d}_{+}^{(2)}\right).
\end{split}
\eeq
The exchange term in this interaction potential can naturally flip an excitation or entangle the particles, being a starting point for a quantum gate scheme~\cite{Ni18}. Here we instead rely on the state-dependent shift of the eigenstates in the presence of an external electric field for the purpose of quantum state engineering. 

We now discuss the spatial geometry of the system, which can be adjusted by changing the trap alignment with respect to the field direction and tuning the distance between the tweezer potential minima. Each optical tweezer is modeled as a cylindrically-symmetric anisotropic harmonic trap with frequencies $\omega_z=\omega$, $\omega_x = \omega_y = \omega_\perp=\eta \omega$ (note that the $z$ direction is chosen parallel to the electric field) with large anisotropy $\eta$. 
One can define here the characteristic trap length scales $a_{\rm ho}=\sqrt{\hbar/\mu\omega_z}$ and $l_\perp=\sqrt{\hbar/\mu\omega_\perp}$, where $\mu$ is the reduced mass of the  system.
For a harmonic trapping potential, the center-of-mass and relative motion are decoupled, and in our analysis we can focus on the relative motion described by the following Hamiltonian:
\beq\label{Htrap}
H_{\rm trap}=-\frac{\hbar^2}{2\mu}\nabla^2+\frac{1}{2}\mu\omega^2 \big{(}\eta^2 x^2+ \eta^2 y^2+(z-a)^2\big{)}\,,
\eeq
where $a$ is the separation between the two traps.
For large anisotropy $\eta \gg 1$ and sufficiently separated molecules, the transverse excitation is the highest energy scale in the system and the motion becomes effectively one-dimensional.
In this regime, one can integrate out perpendicular degrees of freedom~\cite{Deuretzbacher2010, DeuretzbacherPRA13}, assuming that the transverse wave function corresponds to the ground state of the harmonic trap.
This yields an effective full one-dimensional Hamiltonian
\begin{equation}\label{Hfull}
H = -\frac{\hbar^2}{2\mu}\frac{\partial^2}{\partial z^2}+\frac{1}{2}\mu\omega^2 (z-a)^2+ V_{\rm dd}^{ \rm eff} \left(\frac{z}{l_\perp}\right) + H_{\rm rot}^{(1)} + H_{\rm rot}^{(2)},
\end{equation}
where
\beq
\begin{split}
V_{\rm dd}^{ \rm eff}(u)=-\frac{d^2}{8l^3_{\perp}}(1+3\cos(2\theta)) \left(-2u + \sqrt{2\pi}(1+u^2)e^{u^2/2}\mathrm{erfc}(u/\sqrt{2})-\frac{8}{3}\delta(u) \right) \left(d_0^{(1)}d_{0}^{(2)}+\frac{1}{2}d_+^{(1)}d_{-}^{(2)}+\frac{1}{2}d_-^{(1)}d_{+}^{(2)}\right)
\end{split}
\eeq
with $u=z/l_\perp = z\eta^{1/2}/a_{\rm ho}$ and $\mathrm{erfc}$ being the error function. Note that the part of the interaction involving internal states remains unchanged in the effective intermolecular interaction.
Here, $H_{\rm rot}^{(k)}$ denotes the internal state Hamiltonian (Eq.~\eqref{Hrot}) of the $k$-th molecule.
In this work we followed the assumption that the transverse wave function is limited to the lowest oscillator mode, but expanding it on excited modes for a less anisotropic system is straightforward as we have checked that also for higher modes the coefficient in front of the delta function remains $8/3$. If necessary, one can also add an additional short-range interaction to the effective potential in order to reproduce some physical scattering length.
\subsection{Diagonalization}

As a convenient basis for diagonalization of the Hamiltonian we take the states $\ket{i} \equiv \ket{n_z, j^{i}_1, m^{i}_1, j^{i}_2,m^{i}_2}$, where $n_z$ denotes the eigenstate of the harmonic oscillator in the $z$~direction and  $j^{i}_{1(2)}, m^{(i)}_{1(2)}$ denote the rotational states of the first (second) molecule. The function $\ket{n_z}$ can be centered either at $z=0$ or at $z=a$. The former choice is better suited to describe the case in which the molecules lie close to each other or even form a bound state, while the latter should work well if the interaction is weak and the particles are well separated. We checked that at the length scales relevant for our case, better numerical stability is achieved using states centered at $z=0$, while at larger separation or weaker interactions, it would be beneficial to use the other basis. 

Let us now briefly discuss the matrix elements $\bra{i}H\ket{i'}$ of the Hamiltonian of Eq.~\eqref{Hfull}. Starting with the trapping potential along the $z$-axis, $H_{\rm trap}$ of Eq.~\eqref{Htrap}, the integral
\begin{equation}
\begin{split}
\bigg{\langle} n \bigg{|} -\frac{\hbar^2}{2\mu}\frac{\partial^2}{\partial z^2}  +\frac{1}{2}\mu \omega^2(z-a)^2 \bigg{|} n' \bigg{\rangle} = \left[\hbar \omega(n + \frac 12) +\frac 12 \mu \omega^2 a^2\right] \delta_{n,n'} - \hbar \omega \frac{a}{\sqrt{2}a_{\rm ho}} (\sqrt{n'+1}\delta_{n,n'+1} + \sqrt{n} \delta_{n,n'-1})
\end{split}
\end{equation}
is calculated analytically, while the spatial part of the interaction potential, $\bra{i} V_{\rm dd}^{\rm eff} \ket{i}$, is calculated numerically.
The matrix element of the rotational angular momentum for the $k$-th molecule are
\begin{equation}
\bra{i} B\oper{j}^{2}_{k}\ket{i'} =  B \delta_{i, i'}j_{k}^i(j_{k}^i+1).
\end{equation}
The effect of the electric field on the $k$-th molecule and the matrix elements of the dipolar interaction can be calculated from the definition of the dipole moment operator~\cite{Brown}. For completeness, we provide here the matrix element of $d_q$ required to calculate both quantities
\beq
\left<j\pm 1,m+q\right|d_q\left|j,m\right>=d\left(j,m;1,q|j\pm 1,m+q\right)\left(j,0;1,0|j\pm 1,0\right)\sqrt{\frac{2j+1}{2(j\pm 1)+1}}\, ,
\eeq
where $(a,m_a;b,m_b|c,m_c)$ are the Clebsch-Gordan coefficients.

\section{Results}\label{sec:results}
\subsection{Energy spectrum}\label{sec:spectrum}
We now discuss the properties of the energy spectrum of the system. For the physically realistic case in which $\omega\ll B$, the eigenstates separate into branches corresponding to different numbers of rotational excitations. As the most intuitive experimental control knobs for the system are the trap separation and electric field magnitude, we study the energy levels as a function of these parameters. For our numerical calculations we use the values corresponding to NaCs molecules with $d=4.607\,$Debye and $B=1.813\,$GHz, taking the trap frequency $\omega=2\pi\cdot 50\,$kHz and $\eta=10$. This implies $a_{\rm ho} \approx 960.7\, a_0$.

\begin{figure*}[t]
	\begin{center}
		\includegraphics[width=0.99\textwidth]{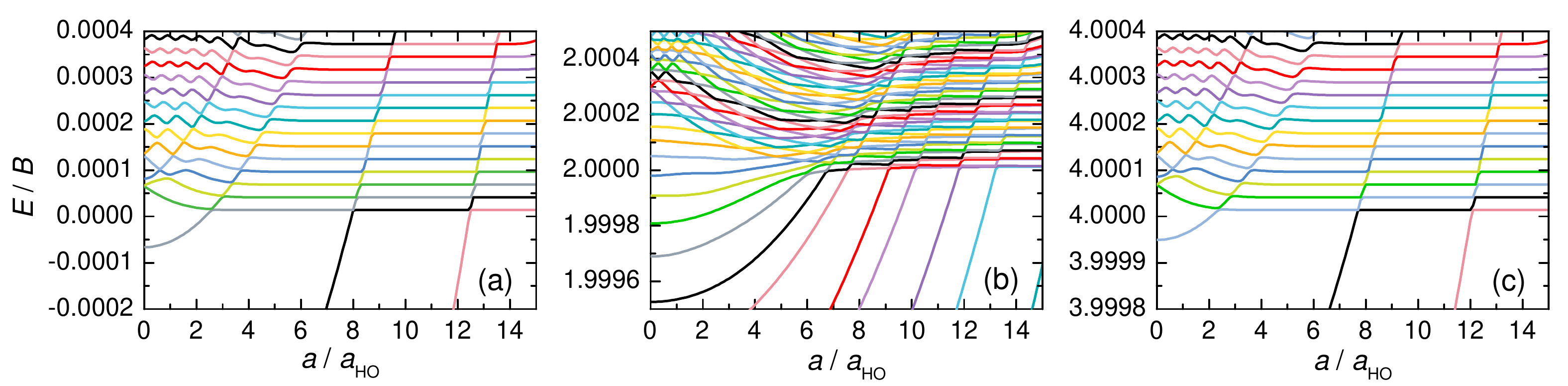}
	\end{center}
	\caption{Energy spectra corresponding to the lowest branch for (a) $M=0$, (b) $M=1$, (c) $M=2$ as a function of the distance between the molecules.  Dipole moment is $d=4.607$ Debye (NaCs) and the electric field $\beta=0$.}
	\label{fig:spectra}
\end{figure*}

First, we investigate the role of trap separation in figure~\ref{fig:spectra} which shows the three lowest branches corresponding to the total angular momentum projections $M=m_1+m_2=0$, 1, and 2. In all cases, one can clearly distinguish two types of eigenstates: the trap states, whose energy is roughly independent of the distance $a$; and the bound states for which the energy goes up roughly as $a^2$. This behavior is typical for the chosen system geometry~\cite{Krych2009} and does not depend on the specific type of interactions. Here, in contrast to the commonly studied contact interaction case, strong attractive potential well leads to the emergence of multiple bound states and large energy shifts. Suitable states for quantum gate realization correspond to the trap states that can be efficiently prepared in remote traps and then brought together. Note that the dipolar bound states display anticrossings with the trap states. This phenomenon, called the trap-induced resonance~\cite{Stock2006}, stems from interaction-induced coupling and in the energy spectrum looks similar to a Feshbach resonance. It can be utilized for different purposes such as production of molecular states, but also to shape the energy spectrum by shifting the energy of a trap state in one of the branches while leaving the other intact. For example, in figure~\ref{fig:spectra}b there are more than five resonances with the lowest trap state, while only two are visible in panel~\ref{fig:spectra}a. At small distances, all types of states become strongly mixed. 

It is important to note that the second branch corresponding to the total angular momentum projection $M=1$ experiences much stronger effects of the dipole-dipole interaction than other branches. It can be understood on the basis of perturbation theory calculations, where the dipole-dipole interaction has a non-zero effect in the first order only for pairs of states with  $j_1=0(1), \, j_2=1(0)$, while for $j_1=j_2=0$ and $j_1=j_2=1$ it contributes only as the second-order correction.

\begin{figure*}
	\includegraphics[width=0.99\textwidth]{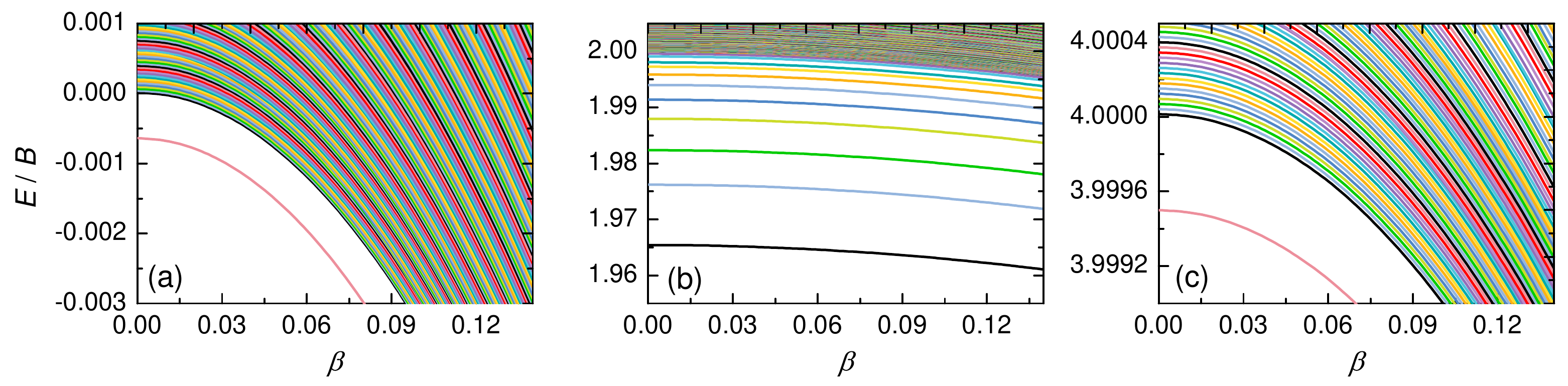}
	\caption{Energy spectra corresponding to the lowest branch for (a) $M=0$, (b) $M=1$, (c) $M=2$ as a function of the electric field $\beta$. Dipole moment is $d=4.607$ Debye (NaCs) and the distance between the molecules is $a/a_{HO}=10.4$.}
	\label{fig:beta}
\end{figure*}

Turning on the electric field induces a nonzero net dipole moment in the molecules and thereby leads to stronger interactions. This is reflected in the spectra as strong shifts of the energy of all two-body states to lower values, as illustrated in figure~\ref{fig:beta}, which shows the energies of the lowest lying trap states in different branches. As we have chosen the field direction to be parallel to the trap axis, the interactions are attractive. Here we chose the trapping potential to be the weakest along the dipole axis as well as the trap alignment, which enhances the attraction and simplifies the spectrum, allowing us to neglect the transverse excitations. In general, it is more convenient experimentally to realize a different setting, while our setup would require an additional light sheet or a lattice. For the general case the density of levels would increase and the spectra would become more complicated, especially the trap-excited states. However, the trap-induced resonances would still be present. 



\subsection{Dynamics}
\label{sec:gate}
Having understood the basic features of the energy spectrum, we now proceed with the dynamics.  We note that recent proposals for quantum engineering protocols involving polar molecules include taking advantage of the dipolar exchange interaction~\cite{Ni18} or utilizing a microwave pulse~\cite{Hughes19}. Our scheme is complementary to these approaches, being based on applying an electric field pulse. The interaction with the field leads to state-dependent energy shifts, allowing in principle for the realization of various quantum gates, but optimization of the pulse shape would be necessary to achieve the desired phase accumulation. Here we will focus our analysis mainly on the simple scenario of electric field quenches and study the population of motionally excited states during the process. While such excitations can be regarded as a fidelity leakage source, it is possible to take them into account and design suitable control pulses that will not only keep the final state close to the trap ground state but also use the full space of states as a resource for gate speedup~\cite{Glaser2015,Muller2021}.

In order to perform a quantum gate, one has first to choose the suitable computational basis. States with varying $M$ are natural candidates for this purpose, as they are not coupled with each other by the dominating interaction term of equation~\eqref{interaction} that we consider here. We thus choose to focus on the two-particle states that have the largest overlap with the pair of molecules being trapped in the motional ground state of separated tweezers. The single qubit states can then be chosen as $|0\rangle=|j=0,\, m=0\rangle$ and $|1\rangle=|j=1,\, m=1\rangle$. Rotational excitations are long-lived and thus very suitable for our purpose. Then the two-qubit basis is composed of states $|00\rangle$ with $J=0$, $M=0$, $|+\rangle$ being the symmetric combination with $J=1$, $M=1$, $|-\rangle$ being the antisymmetric state from the same branch and $|11\rangle$ with $J=2$, $M=2$. In each case we assume the state is initially prepared in a motional eigenstate with the corresponding amplitude $c_i$ ($i$ for initially occupied). The evolution can lead to occupation of multiple other trap states which we will denote with the $c_f$ coefficient ($f$ for final).

We start the calculation by diagonalizing the Hamiltonian for a zero electric field, as in our scenario of interest the field strength is the only parameter that varies during the process. This solution has the advantage of being conceptually simple, while providing short operation times. Local manipulation of the qubits can be achieved by individually addressing the molecules e.g. with an off-resonant laser. Then we move to the interaction picture and solve the corresponding Schr\"{o}dinger equation numerically (using the solver implemented in {\it Mathematica}) for the given time dependence of $\beta(t)$. As before, we choose NaCs molecules as an example and set the distance between the traps to $10.4\,a_{\rm ho}$, which for the system parameters we consider equals~$10^4\,a_0$, away from the trap-induced resonances in the $M=0$ and $M=2$ branches visible in figure~\ref{fig:spectra}. 

\subsubsection{Time evolution with constant electric field}
We first study the quench scenario in which the electric field is suddenly turned on and remains constant throughout the evolution. The operation time is set to $200\,$ns. We will see that this time is long enough for multiple oscillations of the wave function components.

\begin{figure*}[t]
	\begin{center}
		\includegraphics[width=0.99\textwidth]{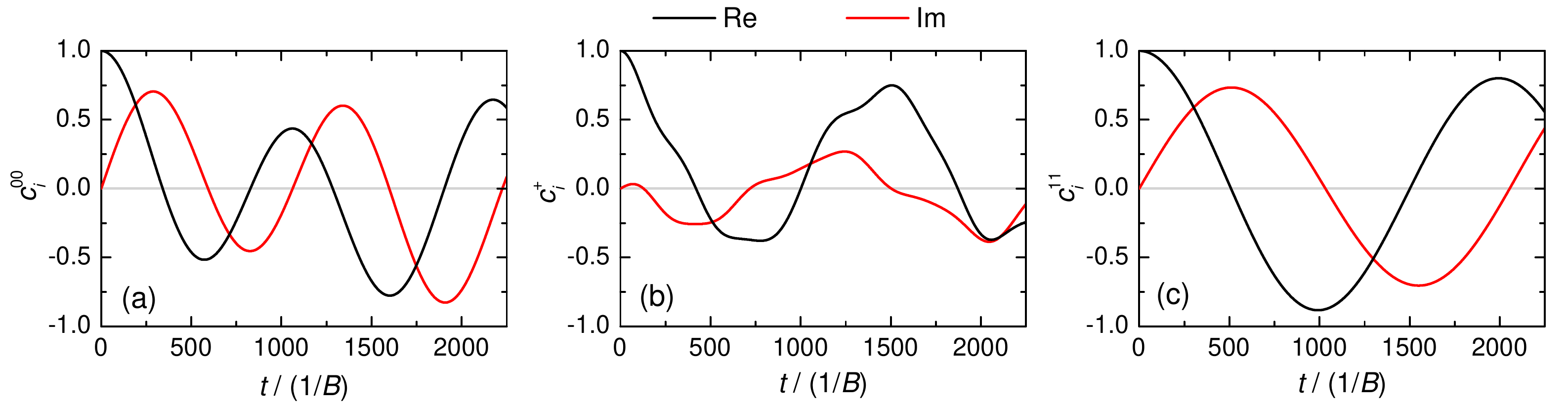}
	\end{center}
	\caption{Time evolution of the three initial states from (a) $M=0$, (b) $M=1$, and (c) $M=2$ branches after switching on a weak electric field $\beta=2.3\cdot 10^{-2}$. Black (red) lines show the real (imaginary) part.}
	\label{fig:weakq}
\end{figure*}

\begin{figure*}[t]
	\begin{center}
		\includegraphics[width=0.99\textwidth]{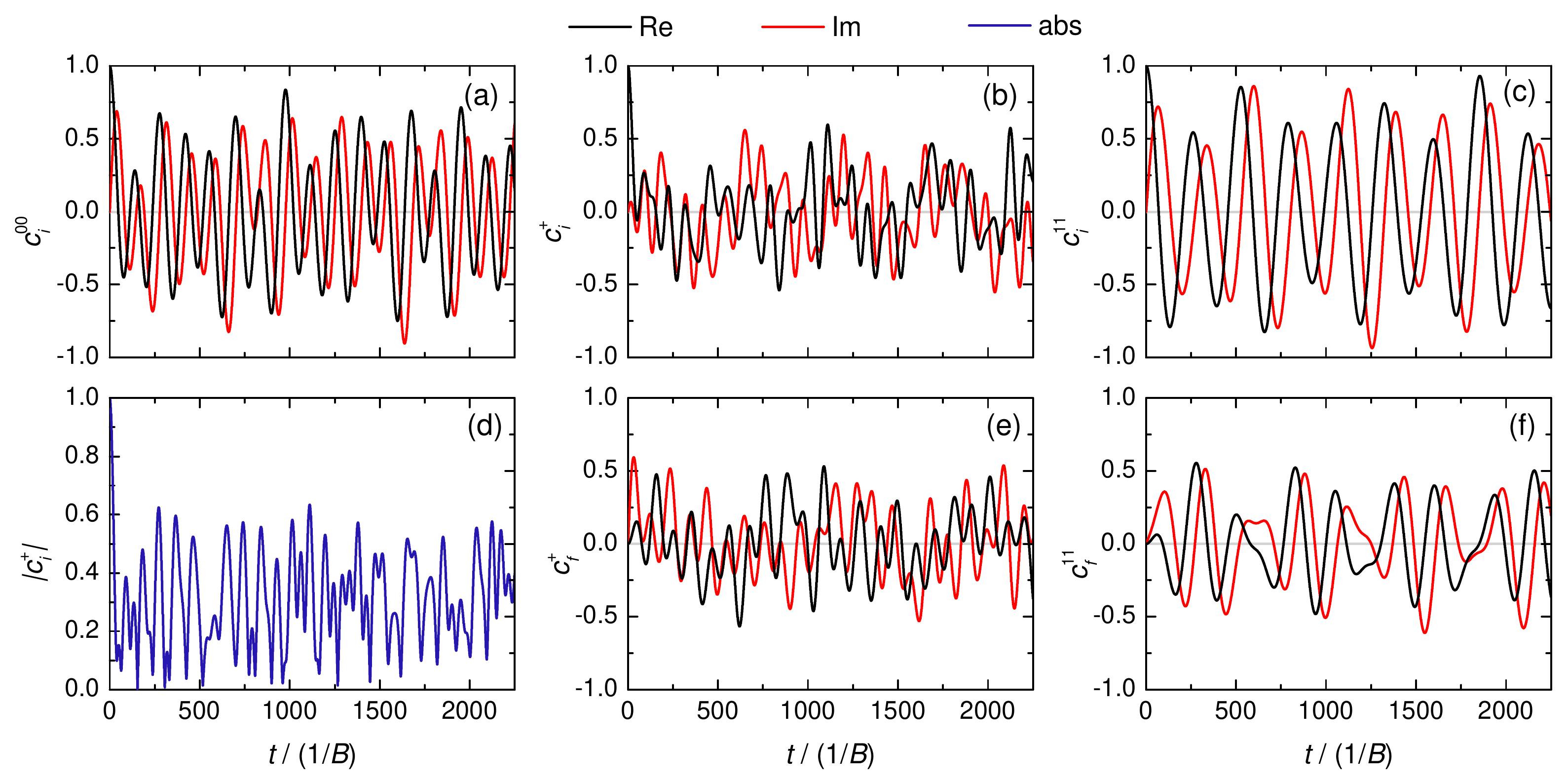}
	\end{center}
	\caption{Upper panel: time evolution of the three states from (a) $M=0$, (b) $M=1$, and (c) $M=2$ branches after switching on a moderate electric field $\beta=1.6\cdot 10^{-1}$. Black (red) lines show the real (imaginary) parts. (d) The absolute occupation of the initial $M=0$ state during the operation. (e) The evolution of an exemplary state from the $M=1$ branch strongly coupled to the initial one. (f) The evolution of an exemplary state from the $M=2$ branch strongly coupled to the initial one.}
	\label{fig:strongq}
\end{figure*}

Let us start with the case of a weak field, shown in figure~\ref{fig:weakq} for $\beta=2.3\cdot 10^{-2}$. The evolution of $M=0$ and $M=2$ states in panels (a) and (c) shows similar oscillatory behavior, while the $M=1$ state in panel (b) undergoes more complicated dynamics. This can be explained by the larger density of states in this branch due to the strong dipolar attraction which mixes trap and bound states. In general, for the electric field values studied here, we observe that the evolution does not lead to coupling of the initially occupied trap state to the bound states and does not excite higher rotational branches. The main couplings occur between the nearby trap states which have the largest overlap with the starting one. For the case depicted in figure~\ref{fig:strongq} the field is increased to $\beta=0.16$. Here the evolution becomes much faster and the population spreads over a larger number of motional basis states. This is once more especially visible for the $M=1$ state, where the population of the initially occupied state drops to below 20\% as shown in panel~\ref{fig:strongq}(d). For reference we also show in panels~\ref{fig:strongq}(e), (f) the occupation of two states that are initially not present, while their population arises due to strong overlap with the initial state after the quench.

\subsubsection{Shaping the pulse}
While the quench scenario is instructive, the ultimate goal would be to switch the field on and off in a smooth way and control its shape in order to achieve desired operations. In order to realize a quantum gate, it is required that the internal levels acquire specific state-dependent phases (e.g. a CNOT operation), while the final motional state should not differ from the initial one. As an initial step towards the full quantum computing proposal, here we study the evolution of the system under a simple pulse $\beta(t)=\beta_0 C(t) \sin\left(\frac{\pi t}{\tau}\right)$ with $\tau$ being the operation time, $\beta_0$ being the pulse amplitude, and $C(t)$ being a correction written as truncated Fourier series $C(t)\propto \sum_i{}A_i \cos(\xi_i t)+B_i \sin(\xi_i t)$ in the spirit of the chopped random basis optimization method~\cite{Caneva2011,Muller2021}. We restrict the total operation time to $150\,$ns for this simple demonstration. For an estimate of the quantum speed limit (the time required for achieving close to unit fidelity), one can look at the inverse of the smallest energy level separation. This is roughly given by the trap frequency, which here corresponds to $50$kHz leading to $\sim 20\,\mu$s gate times.

\begin{figure}
	\begin{center}
	\includegraphics[width=0.99\textwidth]{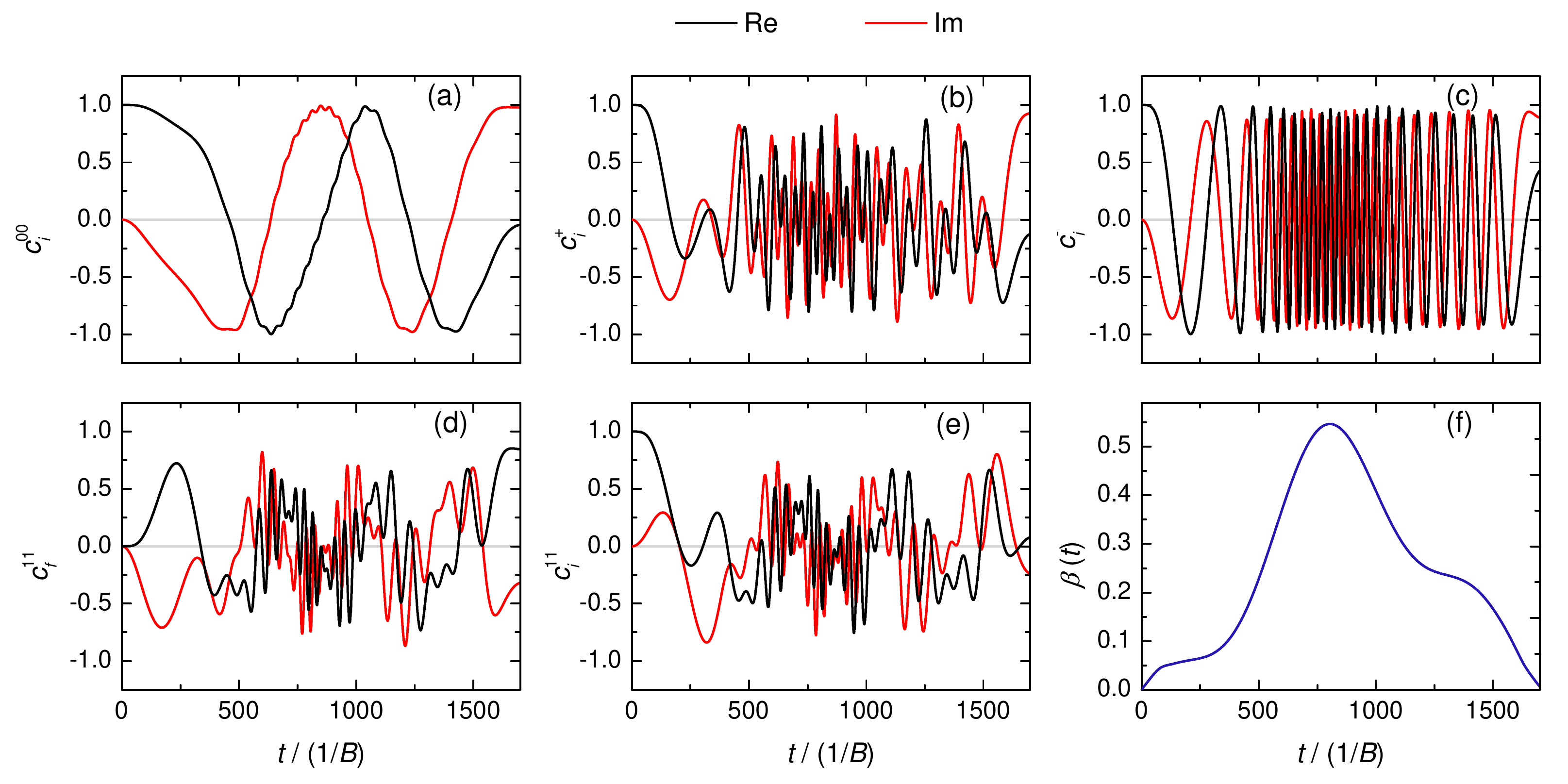}
	\end{center}
	\caption{Time evolution of the qubit states: (a) $|00\rangle$ with $M=0$ , (b) $|+\rangle$ with $M=1$, (c)  $|-\rangle$ with $M=1$, (d) the excited state of $|11\rangle$ with $M=2$, and (e) the initial $|11\rangle$ motional state.  Black (red) lines show the real (imaginary) part. (f) the shape of the electric field pulse used for the evolution.}
	\label{fig:sin}
\end{figure}

 The target of the operation that we chose for the demonstration was to perform a controlled phase gate (in the present case realized in such a way that three states acquire a $\pi$ phase, while the $|11\rangle$ state does not) in the qubit space with only a few optimization steps using gradient search in a small basis set. The system evolution for this case is shown in figure~\ref{fig:sin}. One can notice that if the trap excitations were disregarded, the pulse would already reach 84\% fidelity after performing only two optimization steps. However, the $|11\rangle$ state corresponding to $M=2$ becomes transferred to an excited motional state as a result of the evolution (see figure \ref{fig:sin}(d) and (e)), such that the actual fidelity is zero. This shows that more elaborate control schemes would need to be applied in a realistic system as the motional decoherence can be a problem (note that neglecting the trap dynamics or assuming a gaussian spatial profile with some finite thermal width disregards this problem). The evolution of the $|00\rangle$ state in panel \ref{fig:sin}(a) is notably slower than the others, as its energy is closest to zero, resulting in low oscillation frequency in the interaction picture. For the higher rotational branches the states chosen as the computational basis are more excited and undergo a complicated evolution. The pulse shape shown in panel \ref{fig:sin}(f) is slowly varying and thus experimentally realistic. More extensive calculations using larger basis sets, more optimization steps, and realistic trapping potentials will follow in future work.

\section{Discussion}
\label{sec:disc}

The numerical examples above indicate that while the dynamics does not lead to couplings between the states with different rotational quantum numbers, the trap states can become mixed during the operation. This would lead to the spreading of the wavepacket and limit the number of possible operations. On the other hand, quantum control techniques can be applied to ensure high fidelity~\cite{Glaser2015}. The main question for the system at hand is not the possibility of realizing quantum gates but rather the quantum speed limit achievable in various settings~\cite{Caneva2009}. In a generic setup of two trapped molecules, neglecting possible experimental imperfections, the most important source of fidelity loss lies in the strong dipolar couplings between the motional states from the same rotational branch. This, in principle, affects any possible gate scenario that relies on the dipolar interaction.

Interestingly, the trap-induced resonance phenomenon shown in figure~\ref{fig:spectra} can be utilized to manipulate the energy spectrum and increase the gap between the chosen qubit state and other trap states. Close to the resonance, the trap acquires some bound state character such that part of the wave function is localized at small interparticle separation, while the other states remain delocalized over the trap wqhich reduces their coupling. Furthermore, the energy shift resulting from the anticrossing will be translated to an additional phase shift of the affected state with respect to another qubit state, providing the possibility for speedup.

\section{Conclusions}
\label{sec:conc}
We have analyzed the prospects for quantum state engineering of ultracold polar molecules trapped in separate optical tweezers and controlled using an external electric field. By taking into account the complete structure of the trap states, we have shown how motional excitations could arise during the evolution. This allows for the implementation of more elaborate control schemes, which in principle would allow for working close to the quantum speed limit set by the harmonic confinement frequency. Reaching this goal can be made easier by utilizing trap-induced resonances, which strongly depend on the internal state of the molecules and thus can be precisely tuned.

Possible extensions of the present work include a study of the resonances in experimentally realistic traps and taking into account more details of the interaction potential to deliver precise predictions on the trap-induced resonance positions. Then, optimal control techniques can be utilized to design fast gate protocols with high fidelity. In addition, we suppose that considering more details of particular molecular species, such as including their hyperfine structure and adding more external fields (e.g., microwave) to optimize the qubit space further, will ultimately lead to a full quantum computation toolbox, as well as allow for more detailed studies of molecular interactions. 

\begin{acknowledgments}
We would like to thank Susanne Yelin for useful discussions. 
K.J. was supported by the National Science Centre of Poland grant 2020/37/B/ST2/00486.
A.D. and M.T. were supported by the Foundation for Polish Science within the Homing and First Team programs co-financed by the European Union under the European Regional Development Fund and the PL-Grid Infrastructure.
M.S. and Z.I. acknowledge support from the Polish National Science Centre Grant No. 2015/17/B/ST2/00592.

A.D. as a part of the QOT ICFO group acknowledges support from ERC AdG NOQIA, State Research Agency AEI (“Severo Ochoa” Center of Excellence CEX2019-000910-S) Plan National FIDEUA PID2019-106901GB-I00 project funded by MCIN/ AEI /10.13039/501100011033, FPI, QUANTERA MAQS PCI2019-111828-2 project funded by MCIN/AEI /10.13039/501100011033, Proyectos de I+D+I “Retos Colaboración” RTC2019-007196-7 project funded by MCIN/AEI /10.13039/501100011033, Fundació Privada Cellex, Fundació Mir-Puig, Generalitat de Catalunya (AGAUR Grant No. 2017 SGR 1341, CERCA program, QuantumCAT \ U16-011424, co-funded by ERDF Operational Program of Catalonia 2014-2020), EU Horizon 2020 FET-OPEN OPTOLogic (Grant No 899794), and the National Science Centre, Poland (Symfonia Grant No. 2016/20/W/ST4/00314), Marie Sk\l odowska-Curie grant STREDCH No 101029393, “La Caixa” Junior Leaders fellowships (ID100010434),  and EU Horizon 2020 under Marie Sk\l odowska-Curie grant agreement No. 847648 (LCF/BQ/PI19/11690013, LCF/BQ/PI20/11760031,  LCF/BQ/PR20/11770012).
\end{acknowledgments}

\subsection*{Data availability}
The data that support the findings of this study are openly available in the Bitbucket repository \cite{OurRepo}.

\normalem
\bibliography{mol_gate_bib}


\end{document}